\documentclass[%
 reprint,
 superscriptaddress,
 nobibnotes,
 amsmath,amssymb,
 aps,
prx,
]{revtex4-2}

\usepackage{graphicx}
\usepackage{dcolumn}
\usepackage{bm}
\usepackage{hyperref}

\usepackage{siunitx}
\usepackage{physics}
\usepackage{mhchem}
\usepackage{chemfig}
\newcommand{\makemypolymerdelims}[7][]{%
  \chemmove{\path (#6) -- node[pos=#4] {$\left[\vrule height#2 depth#3 width0pt\hspace{#5}\right]_#1$} (#7);}%
}
\usepackage{float}

\begin{document}


\title{Machine Learning-Assisted Profiling of Ladder Polymer Structure using Scattering}
 
\author{Lijie Ding}
\affiliation{Neutron Scattering Division, Oak Ridge National Laboratory, Oak Ridge, TN 37831, USA}
\author{Chi-Huan Tung}
\affiliation{Neutron Scattering Division, Oak Ridge National Laboratory, Oak Ridge, TN 37831, USA}
\author{Zhiqiang Cao}
\affiliation{Neutron Scattering Division, Oak Ridge National Laboratory, Oak Ridge, TN 37831, USA}
\author{Zekun Ye}
\affiliation{Department of Chemistry, Stanford University, Stanford, CA 94305, USA}
\author{Xiaodan Gu}
\affiliation{School of Polymer Science and Engineering, Center for Optoelectronic Materials and Devices, The University of Southern Mississippi, Hattiesburg, MS 39406, USA}
\author{Yan Xia}
\affiliation{Department of Chemistry, Stanford University, Stanford, CA 94305, USA}
\author{Wei-Ren Chen}
\affiliation{Neutron Scattering Division, Oak Ridge National Laboratory, Oak Ridge, TN 37831, USA}
\author{Changwoo Do}
\email{doc1@ornl.gov}
\affiliation{Neutron Scattering Division, Oak Ridge National Laboratory, Oak Ridge, TN 37831, USA}

\date{\today}

\begin{abstract}
Ladder polymers, known for their rigid, ladder-like structures, exhibit exceptional thermal stability and mechanical strength, positioning them as candidates for advanced applications. However, accurately determining their structure from solution scattering remains a challenge. Their chain conformation is largely governed by the intrinsic orientational properties of the monomers and their relative orientations, leading to a bimodal distribution of bending angles, unlike conventional polymer chains whose bending angles follow a unimodal Gaussian distribution. Meanwhile, traditional scattering models for polymer chains do not account for these unique structural features. This work introduces a novel approach that integrates machine learning with Monte Carlo simulations to address this challenge. We first develop a Monte Carlo simulation for sampling the configuration space of ladder polymers, where each monomer is modeled as a biaxial segment. Then, we establish a machine learning-assisted scattering analysis framework based on Gaussian Process Regression. Finally, we conduct small-angle neutron scattering experiments on a ladder polymer solution to apply our approach. Our method uncovers structural details of ladder polymers that conventional methods fail to capture.
\end{abstract}

\maketitle

\section{Introduction}

Ladder polymers\cite{xia2023ladder, teo2017synthesis, overberger2006ladder,zhou2008promising,scherf1999ladder}, characterized by their rigid, double-stranded, ladder-like structure, have garnered significant interest due to their exceptional thermal stability and mechanical strength. These unique properties make them highly suitable for advanced applications\cite{lai2019tuning,lai2022hydrocarbon,lee2017fully,mckeown2010exploitation} such as electronic devices, membranes, and high-performance materials. It becomes important to characterize the structure and chain conformation of the ladder polymer as its precise molecular arrangement directly affects its physical properties, performance, and suitability for specific applications, allowing for better control over fabrication and optimization for targeted uses. 

Small angle scattering experiment\cite{lindner2002neutrons}, including X-ray scattering\cite{chu2001small} and neutron scattering\cite{chen1986small,shibayama2011small} are often used to study the characteristics of polymer system, and to unveil the single polymer structure using dilute polymer solutions. The scattering data is often analyzed using various polymer models to extract the polymer parameters, e.g. contour length, radius of gyration and persistence length. However, traditional polymer models, such as Gaussian coils\cite{debye1947molecular} or worm-like chains\cite{pedersen1996scattering} are inadequate for capturing the distinctive features of ladder polymers since they are designed to model the single-stranded polymers and discard twisting. These models do not fully represent the inherent rigidity and extended conformation of the ladder polymer, thus fail to provide an accurate depiction of the ladder polymers structure.

To overcome these challenges and provides an accurate description of the ladder polymer structure using scattering data, we build a new model for the ladder polymer, accounting for the biaxial nature of it's monomer structure and inherent rigidity. Due to the complexity of this model, it is difficult to derive the analytical form of the scattering function, which is typically required for fitting scattering data using traditional approaches. To address this, we leverage the power of Machine Learning (ML)\cite{murphy2012machine} and Monte Carlo (MC)\cite{krauth2006statistical} simulations. 

The recent advancements in ML have enabled numerous applications in materials science, including the analysis of scattering data\cite{chang2022machine} without knowing an explicit analytical form. This approach relies on large data sets that include scattering functions and corresponding polymer parameters, allowing ML to learn the relationship between them. Meanwhile, MC can be used to build such data sets. Given a set of polymer parameters, such as contour length and bending rigidity, we can use MC simulation to generate an ensemble of the polymer conformations and calculate the form factor, or scattering function. This combination of ML and MC provides a powerful framework for analyzing complex polymer systems and has been proven useful for various single-stranded polymer systems\cite{tung2022small, ding2024machine}.

In this paper, we present a framework for analyzing the scattering data of ladder polymer using ML. We firstly introduce a model of the ladder polymer where the biaxiality, inherent rigidity and arrangement of successive monomers all play crucial role in determining the polymer conformation. We then carry out MC simulation to generate a large data set of the scattering data and train a ML model of Gaussian process regression\cite{williams2006gaussian} (GPR) to obtain the mapping between scattering data and polymer parameters. Finally, we synthesize ladder polymer samples and measure the scattering function using small-angle neutron scattering (SANS) experiment and apply out method to the extract important polymer parameters for the measured sample.

\section{Model}
To capture the ladder shape and biaxiality of the polymer, we model each monomer unit of the polymer as a rectangular segment whose orientation is specified by two unit vectors, $\vu{u}$ and $\vu{v}$, where $\vu{u}$ is along the along axis of the segment, or the polymer tangent direction, and $\vu{v}$ is along the segment short axis and perpendicular to $\vu{u}$. A polymer is then modeled as a chain of $L$ segments, where $L$ is the contour length in unit of monomer length $B$. 

\begin{figure}[h]
    \centering
    \includegraphics{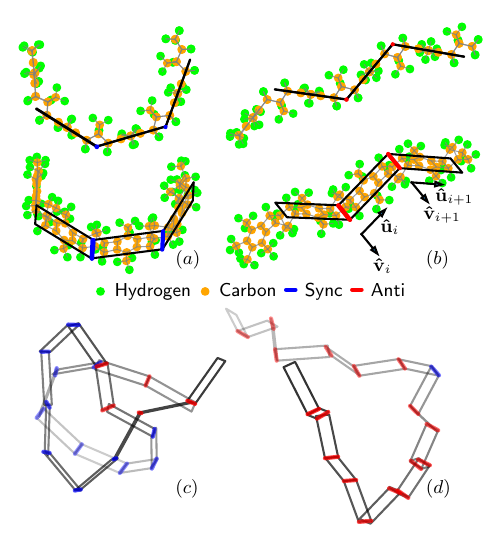}
    \caption{Illustration of the segmentation model of CANAL ladder polymer. (a) Molecular structure of monomer segments connected through Sync link, overlapped with rectangle used in our model, top and bottom two are the same polymer with different point of view. (b) similar to (a), but with segments connected through Anti link. (c) Monte Carlo generated polymer with low anti rate $R_a=0.1$ and (d) high anti rate $R_a=0.9$.}
    \label{fig: model_demo}
\end{figure}

Unlike conventional polymer, the successive segments on the ladder polymer, i.e. Catalytic arene-norbornene annulation (CANAL) polymer, tend to form a angle, as shown in Fig.~\ref{fig: model_demo}(a) and (b). We introduce another two unit vectors, $\vu{u}'$ and $\vu{v}'$, to represent this preferred orientation for the successive segment. For two connecting segments $i$ and $j$, the angle between $(\vu{u}, \vu{v})$ and $(\vu{u}', \vu{v}')$ is the inherent bending and twisting. For this specific model we are concerned of, the $\vu{v}=\vu{v}'$, and we denote the inherent bending $\cos(\alpha) = \vu{u}\cdot\vu{u}'$. There is a energy cost when $(\vu{u}_{i+1},\vu{v}_{i+1})$ tilt away from $(\vu{u}'_i,\vu{v}'_i)$, given polymer energy $E = \sum_{i}\frac{1}{2}K_t\phi_{i}^2 + \frac{1}{2}K_b\theta_{i}^2$, where the bending is $\cos(\theta_{i})= \vu{u}'_i\cdot\vu{u}_{i+1}$, twisting is $\cos(\phi_{i})= \vu{v}'_i\cdot\vu{v}_{i+1}$, and $K_t$ and $K_b$ are the twisting and bending modulus, respectively. 

Finally, the preferred orientation for successive segment at each segment may not stay on the same side. Comparing Fig.~\ref{fig: model_demo}(a) and (b), when they stay on the same side, we call them connected by Sync links, the polymer rolls up and become coil shape as shown in Fig.~\ref{fig: model_demo}(c). On the contrary, if they flip side, or connected by Anti links, the polymer tend to extend longer, as shown in Fig.~\ref{fig: model_demo}(d). We define the probability of a link being a Anti link as anti rate $R_a$.

Given a contour length $L$, inherent bending angle $\alpha$, anti rate $R_a$, bending modulus $K_t$ and twisting modulus $K_b$, the ensemble of ladder polymer configuration is determined. The configuration can be captured by the form factor, given by\cite{chen1986small,lindner2002neutrons}:
\begin{equation}
    \label{equ: S(QB)}
    S(QB) = \frac{1}{L^2}\sum_{i=1}^L\sum_{j=1}^L \frac{\sin(Q |\va{r}_i-\va{r}_j|)}{Q |\va{r}_i-\va{r}_j|}
\end{equation}
where $Q$ is the scattering vector and $\va{r}_i$ is the position vector of segment $i$ and $\va{r}_{i+1} = \va{r}_i + B\vu{u}_i$. In addition, we also calculate the radius of gyration $R_g^2 = \frac{1}{2}\left<(\va{r}_i - \va{r}_j )^2\right>_{i,j}$, with $\left<\cdots \right>_{i,j}$ denoting the average over all pairs of segments. We will use MC and ML to understand the relationship between form factor $S(QB)$ and other polymer parameters $(R_a,\alpha,L,R_g^2,K_t,K_b)$.

\section{Method}

\subsection{Synthesis of CANAL ladder polymer}
\begin{figure*}[!th]
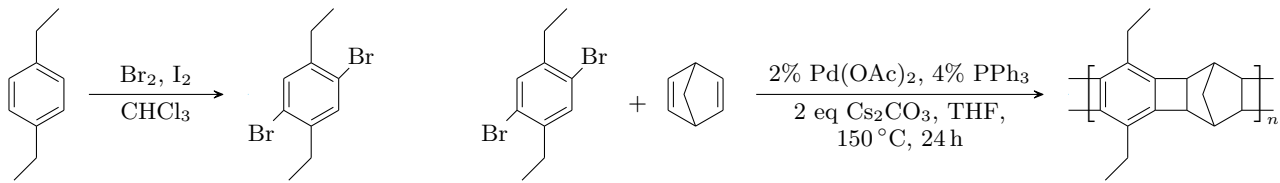

    \centering
    \schemestart
    \chemfig[atom sep=12pt]{*6(-(-[6]-[5])=-=(-[2]-[1])-=)}
    \arrow{->[\ce{Br2}, \ce{I2}][\ce{CHCl3}]}[0, 1.5] 
    \chemfig[atom sep=12pt]{*6(-(-[6]-[5])=-(-[1]Br)=(-[2]-[1])-=)-[5]Br}
    \hspace{1.2cm}
    \chemfig[atom sep=12pt]{*6(-(-[6]-[5])=-(-[1]Br)=(-[2]-[1])-=)-[5]Br}
    \+ 
    \chemfig[atom sep=12pt]{*6(-@{a1}-=-@{b1}-=)}
    \chemmove{\draw[-] (a1) -- +(110:0.45) -- (b1);}
    \arrow{->[2\% \ce{Pd(OAc)2}, 4\% \ce{PPh3}][\parbox{5cm}{\centering 2 eq \ce{Cs2CO3}, THF, \\ \SI{150}{\degreeCelsius}, \SI{24}{h}}]}[0, 2.75]
    \chemfig[atom sep=12pt]{@{L}-[0]([:30]*6(-[{-30}](-[6]-[5])=*4(-*6(-@{a}-([0]-@{R})-([0]-)-@{b}-)--)-=(-[2]-[1])-([4]-)=))}
    \makemypolymerdelims[n]{8pt}{8pt}{.55,yshift=4pt}{60pt}{L}{R}
    \chemmove{\draw[-] (a) -- +(110:0.45) -- (b);}
    \schemestop
    \caption{Illustration of synthesis procedure of the CANAL ladder polymer}
    \label{fig: synthesis}
\end{figure*}

To a flame-dried \SI{15}{mL} glass pressure tube was added 1,4-dibromo-2,5-diethylbenzene\cite{sugamata2021gas} (\SI{584}{mg}, \SI{2}{\mmol}), \ce{Pd(OAc)2} (\SI{9}{mg}, \SI{0.04}{\mmol}), \ce{PPh3} (\SI{21}{mg}, \SI{0.08}{\mmol}) and butylated hydroxytoluene (\SI{1}{mg}). The tube was transferred into a nitrogen-filled glove box, and norbornadiene (\SI{220}{\micro\liter}, \SI{2.2}{\mmol}), \ce{Cs2CO3} (\SI{1.3}{mg}, \SI{4}{\mmol}) and THF (\SI{2}{mL}) was added. The tube was sealed with a Teflon cap and removed from the glovebox. The reaction mixture was heated to \SI{150}{\celsius} for \SI{24}{\hour}. The mixture was then cooled to room temperature and passed through Celite to remove inorganic salts. Chloroform (3*\SI{5}{mL}) was used to wash the residue. The filtrate was concentrated and dissolved in a minimum amount of chloroform, which was then precipitated into methanol. The precipitated polymer was collected by centrifugation, washed with methanol, and dried under vacuum. Fig.~\ref{fig: synthesis} illustrate the synthesis procedure. The obtained polymer were fractionated using Soxhlet Extractor to generate low molecular weight polymer fraction (washed down from ethyl acetate) and high molecular weight polymer fraction (washed down from choloroform)

\subsection{small-angle neutron scattering}
The extended Q-range small-angle neutron diffractometer (EQ-SANS) at the Spallation Neutron Source at the Oak Ridge National Laboratory was used to characterize the conformation of the ladder polymer.\cite{Zhao2010a, Heller2018} Low molecular weight CANAL ladder polymer was dissolved in deuterated 1,2-Dichlorobenzene at 5mg/mL. Two sample-to-detector distances (2.5 m and 4 m) were used with two wavelength bands defined by the minimum wavelength of $\lambda_{min} =2.5 \rm{\AA}$  and $\lambda_{min} = 10 \rm{\AA}$, respectively, to cover scattering wave vectors ranging from 0.006 to 0.5 $\rm{\AA}^{-1}$. The choppers were operated at 60 Hz. The ladder polymer solution was loaded in the quartz cell of 2 mm path length and measured at \SI{25}{\celsius}, \SI{75}{\celsius} and \SI{125}{\celsius}. The measured data were corrected by detector sensitivity and background scattering from the empty cell and then converted into absolute scale intensities ($\rm{cm^{-1}}$) using a porous silica standard sample.\cite{arnold2014mantid, Heller2022} Finally, scattering from the solvent was subtracted before the data is scaled into the unit-less $QB$ axis using the monomer length $B$.

\subsection{Monte Carlo simulation}
To calculate the form factor of the polymer ensemble at various polymer parameters, we sample the configuration space of the polymer using direct sampling\cite{krauth2006statistical}. For a given set of $(R_a,\alpha,L,K_t,K_b)$, we generate 2,000 polymer configurations and calculate the averaged form factor $S(QB)$. The polymer configuration is determined by the link type $l_i$, relative bending and twisting angles $\{(l_i, \theta_i,\phi_i)\}$, letting $l_i=0$ represent Sync link and $l_i=1$ denotes Anti link, the $l_i$ follows a Bernoulli distribution with probability $P(l_i=1)=R_a$. In addition $\theta_i$ and $\phi_i$ follows the Gaussian distribution $\theta_i\sim N(0,\sqrt{K_b/k_B T})$ and $\phi_i\sim N(0,\sqrt{K_t/k_B T})$, as they are independent in the polymer energy $E = \sum_{i}\frac{1}{2}K_t\phi_{i}^2 + \frac{1}{2}K_b\theta_{i}^2$, which follow the Boltzmann distribution $P(E)\sim e^{-E/k_B T}$. After sampling $\{(l_i, \theta_i,\phi_i)\}$ for all segments based on their distribution, we calculate $(\vb{u}_i,\vb{v}_i)$ and $\va{r}_{i+1} = \va{r}_i + B\vu{u}_i$ of each polymer segments, then check the self-avoidance criteria $|\va{r}_i-\va{r}_j|<0.5B$ for all pairs of segments, only configurations satisfying these criteria are kept.

\subsection{Gaussian process regression}
Under the framework of GPR\cite{williams2006gaussian}, the goal is to obtain the posterior $p(\vb{Y}_*|\vb{X}_*,\vb{X},\vb{Y})$ of the function output $\vb{y}$, where $X=\left\{\ln{S(QB)_{train}}\right\}$, $X_*=\left\{\ln{S(QB)_{test}}\right\}$ are the training set and test set, $\vb{Y}$ and $\vb{Y_*}$ are the corresponding polymer parameters ${(R_a,\alpha,L,R_g^2)}$. In our case, we use $70\%$ of the data set $\vb{F}=\left\{\ln{S(QB)}\right\}$ as the training set, and the rest $30\%$ as the test set. The joint distribution is for a Gaussian process is given by \eqref{equ: Gaussian_process}

\begin{equation}
    \mqty(\vb{Y}~\\ \vb{Y}_*) \sim \mathcal{N}\left( \mqty[m( \vb{X}~) \\ m(\vb{X}_*)], \mqty[k(\vb{X}~,\vb{X}~) & k(\vb{X}~,\vb{X}_*) \\ k(\vb{X}_*,\vb{X}~)~ & k(\vb{X}_*,\vb{X}_*)]  \right)
    \label{equ: Gaussian_process}
\end{equation}

where a constant prior mean $m(\vb{x})$ and a linear combination of a Radial basis function (Gaussian) kernel and white noise for the kernel $k(\vb{x},\vb{x}') = e^{\frac{-|\vb{x} - \vb{x}'|^2}{2l}} + \sigma \delta(\vb{x},\vb{x}')$ are used, in which $l$ is the correlation length, $\sigma$ is the variance of observational noise and $\delta$ is the Kronecker delta function.

\section{Results}
We prepare the data set $\vb{F}=\left\{ \ln{S(QB)} \right\}$ by generating conformations of ladder polymers using MC for 6,000 random combination of $(R_a,\alpha,L,K_t,K_b)$ and calculate the corresponding $R_g^2$ and $S(QB)$. The $S(QB)$ are calculated for 100 different $QB\in[0.07,3]$, such that the $\ln{QB}$ grid is uniformly placed in this interval. The polymer parameters are sampled as $R_a\sim U(0,1)$, $\alpha\sim\frac{\pi}{180}U(45, 60)$, $L\sim U(4,50)$, $K_t\sim U(50,100)$ and $K_b\sim U(50,100)$, where $U(a,b)$ is the uniform distribution in interval $[a,b]$. Natural units are used, such that length are in unit of segment or monomer length $B$, and energy are measured in unit of thermal noise $k_B T$. We firstly study the effect of polymer parameters on the form factor, then validate the feasibility for ML inversion of each polymer parameter, train a GPR and test it using MC generated data. Finally, we carry out SANS experiment and applied the trained GPR to the experimentally obtained form factor.

\subsection{Form factor of the ladder polymer}
\begin{figure}[!h]
    \centering
    \includegraphics{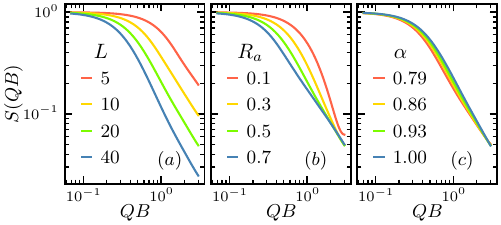}
    \caption{Examples of simulated form factor $S(QB)$ versus scattering vector $Q$ normalized by monomer length $B$, with $K_t=K_b=100$ at various contour length $L$, anti rate $R_a$ and inherent bending angle $\alpha$. (a) $S(QB)$ at various $L$ with $R_a=0.5$ and $\alpha=0.93$ (or $53.3^\circ$) (b) $S(QB)$ at various $R_a$ with $L=20$ and $\alpha=0.93$. (c) $S(QB)$ at various $\alpha$ with $L=20$ and $R_a=0.5$.}
    \label{fig: S(QB)}
\end{figure}
While the polymer energy is only directly related to the bending modulus $K_b$ and twisting modulus $K_t$, these modulus are relative large as the segments are connected by strong chemical bonds, leaving the major conformation change determined by the inherent bending angle $\alpha$, and anti rate $R_a$. These conformation change is captured by the form factor. Fig.~\ref{fig: S(QB)} shows the $S(QB)$ at various contour length $L$, anti rate $R_a$ and inherent bending angle $\alpha$. As shown in Fig.~\ref{fig: S(QB)}(a), increasing the contour length lead to rapid decrease of $S(QB)$, resulting from the extension of the polymer that increase the scattering at low $Q$. Fig.~\ref{fig: S(QB)}(b) shows that increasing anti rate $R_a$ has similar effect of increasing $L$, as it also make the polymer extending longer. Increasing inherent bending angle $\alpha$ make the opposite effect and increase the $S(QB)$ as it effectively make the polymer more straight.

\subsection{Feasibility of Machine Learning inversion}

To access the feasibility of using GPR to map the form factor $\vb{F}=\{\ln{S(QB)}\}$ to polymer parameters $\vb{Y}=\{(R_a,\alpha,L,R_g^2,K_t,K_b)\}$, following the similar ML inversion framework\cite{chang2022machine}, we carry out principle component analysis of $6,000\times100$ matrix $\vb{F}$, by decomposing it into $\vb{F}=\vb{U}\vb{\Sigma}\vb{V}^T$ using singular value decomposition (SVD), where $\vb{U}$, $\vb{\Sigma}$, and $\vb{V}$ are matrices of $6,000\times6,000$, $6,000\times100$, and  $100\times100$ sizes, respectively. $\vb{V}$ is consist of the singular vectors, and the entries of $\vb{\Sigma}^2$ are proportional to the variance of the projection of $\vb{F}$ onto corresponding principal vectors in $\vb{V}$.

\begin{figure}[hbt]
    \centering
    \includegraphics[width=1\linewidth]{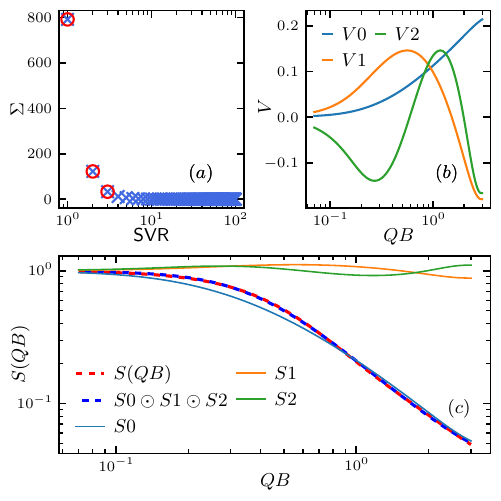}
    \caption{Singular value decomposition (SVD) of scattering function data set $\vb{F} = \left\{\ln S(QB)\right\}$. (a) Singular value $\Sigma$ versus Singular Value Rank (SVR), value with top 3 rank are highlighted in red circle. (b) First 3 singular vectors $V1,V2$ and $V3$. (c) Decomposition of $\ln{S(QB)}$ with $L=20$, $\alpha=0.93$, $R_a=0.5$ and $K_b=K_t=100$, $\ln{S0},\ln{S1}$ and $\ln{S2}$ are the projection of $\ln{S(QB)}$ onto $V0, V1$ and $V2$, respectively, e.g. $S0(QB) = \exp{V0(QB)\sum_{Q'}\ln{S(Q'B)}V0(Q'B)}$, and $\odot$ denotes the Hadamard, or entrywise, product, i.e. $(a\odot b)_i = a_i b_i$.}
    \label{fig: SVD_Sq}
\end{figure}

As shown in Fig.~\ref{fig: SVD_Sq}(a), the singular value decays rapidly with its rank, suggesting the projecting $\ln{S(QB)}\in\vb{F}$ onto the space spanned by the high rank singular vectors manifest good approximation of the entire $\ln{S(QB)}$. Fig.~\ref{fig: SVD_Sq}(b) shows the the first three singular vectors $(V1,V2,V3)$, and Fig.~\ref{fig: SVD_Sq}(c) demonstrate the projection of $\ln{S(QB)}$ on to these top 3 singular vectors do recover the original $\ln{S(QB)}$ very well.

\begin{figure}[hbtp]
    \centering
    \includegraphics[width=1\linewidth]{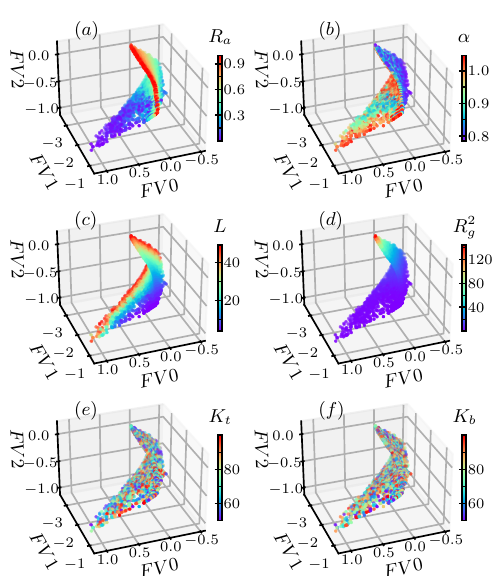}
    \caption{Distribution of various inversion targets of data set $\vb{F} = \left\{\ln S(QB)\right\}$ projected into the singular value space $(FV0,FV1,FV2)$ described by the first 3 singular vectors $(V0,V1,V2)$. (a) Anti rate $R_a$. (b) Inherent bending angle $\alpha$. (c) Contour length $L$. (d) Radius of gyration square $R_g^2$. (e) Twisting modulus $K_t$ and (f) Bending modulus $K_b$.}
    \label{fig: SVD_feature}
\end{figure}

By projecting the $\vb{F}=\left\{\ln{S(QB}\right\}$ onto the singular vector space of $(V0,V1,V2)$, each $\ln{S(QB)}$ become a coordinate in the three dimensional space, $(FV0,FV1,FV2)$, and the entire set of coordinates provides a good proxy of the raw data set $\vb{F}$. By plotting the distribution of polymer parameters $\vb{Y}$ in the $(FV0,FV1,FV2)$, Fig.~\ref{fig: SVD_feature} provide insight for the feasibility of ML inversion of each of the polymer parameter, int which the corresponding value are represented by color distribution. 

As shown in Fig.~\ref{fig: SVD_feature}(a)-(d), the polymer parameters $(R_a,\alpha,L,R_g^2)$ are well spread out in the $(FV0,FV1,FV2)$ space,indicating a good reversed mapping from $\ln{S(QB)}$ to these parameters, indicating they are good inversion targets. On the contrary, Fig.~\ref{fig: SVD_feature}(e) and (f) show that the distribution of the bending and twisting modulus $K_b$ and $K_t$ are rather random, suggesting there they can not be easily extract from the $\ln{S(QB)}$. This is in line with our expectation as the conformation of the ladder polymer is not sensitive to the wiggling around the inherent bending angle $\alpha$ since $\alpha$ is very large compare to the flexibility of the chemical bond.

\subsection{Machine Learning inversion of simulation data}

\begin{figure}[hbtp]
    \centering
    \includegraphics{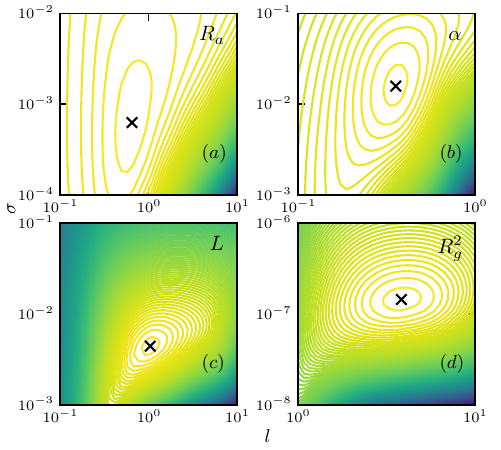}
    \caption{Log marginal likelihood surface of hyperparameters $l$ and $\sigma$ for various polymer parameters, with optimized value marked with black cross. (a) Anti rate $R_a$. (b) Inherent bending angle $\alpha$. (c) Contour length $L$. (d) Radius of gyration $R_g^2$.}
    \label{fig: LML}
\end{figure}

With the feasibility for ML inversion for $(R_a,\alpha,L,R_g^2)$ from the $\ln{S(QB)}$ established, we test such inversion using simulation data. We divide the data set $\vb{F}=\left\{\ln{S(QB)}\right\}$ into two parts, a training set $\left\{ \ln{S(QB)_{train}}\right\}$ consisting $70\%$ of $\vb{F}$ and a testing set $\left\{ \ln{S(QB)_{test}}\right\}$ made of the rest $30\%$. We optimize the hyperparameters of the GPR model using the training set for each polymer parameter and then extract the corresponding polymer parameters $(R_a,\alpha,L,R_g^2)$ from the $\ln{S(QB)} \in \left\{ \ln{S(QB)_{test}}\right\}$. The scikit-learn Gaussian Process library\cite{pedregosa2011scikit} was used for the training. Tab.~\ref{tab: hyperparameter} shows the optimized hyperparameters for each polymer parameters, obtained by maximizing the log marginal likelihood\cite{williams2006gaussian} as shown in Fig.~\ref{fig: LML}.

\begin{table}[hbtp]
    \centering
    \begin{tabular}{|*{3}{p{0.25\linewidth}|}}
        \hline
                 & $l$ & $\sigma$ \\ \hline
        $R_f$ &  \num{6.497e-01} & \num{6.301e-04} \\ \hline  
        $\alpha$ &  \num{3.570e-01} & \num{1.585e-02}  \\ \hline      
        $L$ & \num{1.043e+00} & \num{4.442e-03} \\ \hline 
        $R_g^2$ & \num{3.843e+00} & \num{1.447e-07} \\ \hline 
    \end{tabular}
    \caption{Optimized hyperparameters for each features, obtained from maximum log marginal likelihood.}
    \label{tab: hyperparameter}
\end{table}

\begin{figure}[hbtp]
    \centering
    \includegraphics{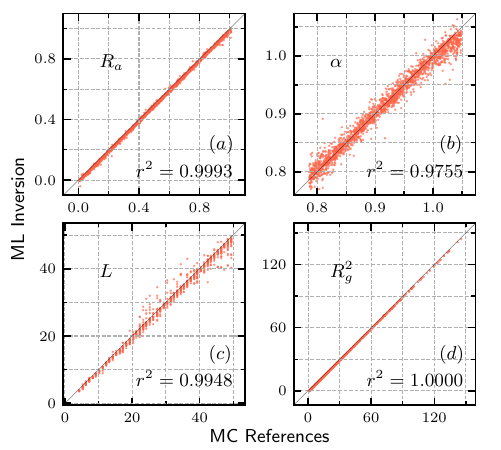}
    \caption{Comparison of polymer parameters in simulation and inverted by machine learning. (a) Anti rate $R_a$. (b) Inherent bending angle $\alpha$. (c) Contour length $L$. (d) Radius of gyration $R_g^2$.}
    \label{fig: ML inversion}
\end{figure}

Fig.~\ref{fig: ML inversion} shows the comparison between the polymer parameters $(R_a,\alpha,L,R_g^2)$ obtained from ML inversion and the corresponding MC references. The data agree very well, and lie closely to the diagonal line, with coefficient of determination $r^2$ score close to 1. The high precision highlights the effectiveness of extracting key parameters from the form factor and further confirms the robustness of our GPR model.

\subsection{Analysis of experimental measurement}
To put our ML inversion model into practice, we synthesize CANAL ladder polymer and carry out small-angle neutron scattering (SANS) experiment to measure it's form factor. 

\begin{figure}[hbtp]
    \centering
    \includegraphics{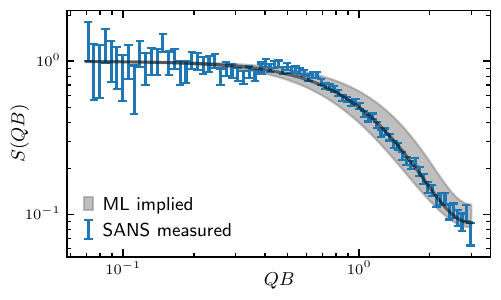}
    \caption{Experimentally measured ladder polymer form factor $S(QB)$ using small-angle neutron scattering with $Q$ normalized by $B=8.12$\rm{\AA}, and MC reconstructed $S(QB)$ based on ML implied polymer parameters. The Dark gray line is calculated using $(L,R_f,\alpha)=(12,0.14,0.89)$, gray region indicate the uncertainty with upper bound calculated using $(L,R_f,\alpha)=(9,0.07,0.94)$, and lower bound using $(L,R_f,\alpha)=(14,0.21,0.84)$.}
    \label{fig: exp_S(QB)}
\end{figure}

Fig.~\ref{fig: exp_S(QB)} shows normalized form factor measured from the SANS experiment and the ML implied curve. The SANS measured $S(QB)$ shows good flat part in the low $Q$ region in the log-log plot, allow us to fit for the normalization coefficient using Guinier approximation\cite{lindner2002neutrons,guinier1955small} $S(QB)\sim e^{-(QR_g)^2/3}$, and the monomer length $B$ obtained by molecular structure optimization allow us to rescale the horizontal axis. By feeding the normalized experimental $\ln{S(QB)}$ to the trained GPR, we obtain the polymer parameters $(R_a,\alpha,L,R_g^2)$, as shown in Tab.~\ref{tab: ML_polymer_parameter}, and then run MC simulation with these parameters to reconstruct the ML implied $S(QB)$. The SANS measured $S(QB)$ and the Ml implied one agree closely.

\begin{table*}[hbtp]
    \centering
    \begin{tabular}{|p{0.3\linewidth}|*{4}{p{0.13\linewidth}|}}
        \hline
                & $R_a$ & $\alpha$ & $L$ & $R_g^2$ \\ \hline
        Machine Learning inversion & $0.14\pm0.07$ & $0.89\pm0.05$ & $11.6\pm2.8$ & $2.07\pm0.28$ \\ \hline  
        Molecular structure optimization$^\dagger$ \cite{shankar2022introduction} & N/A & $0.96\pm0.08$ &  N/A  & N/A  \\ \hline 
        Flexible cylinder fitting\cite{chen2006incorporating,pedersen1996scattering} & N/A & N/A  & $12.0\pm0.8$ & N/A  \\ \hline 
        Guinier approximation fitting\cite{lindner2002neutrons,guinier1955small} &  N/A & N/A & N/A & $1.82\pm0.15$  \\ \hline      
    \end{tabular}
    \caption{Comparison of ladder polymer structure parameters extracted from scattering function using Machine Learning and other traditional methods. ($^\dagger$The atomistic structure of the ladder polymer with 4 monomer units were optimized using the Forcite Module with COMPASS force field in Materials Studio 8.0, BIOVIA.) }
    \label{tab: ML_polymer_parameter}
\end{table*}

\begin{figure}[hbtp]
    \centering
    \includegraphics{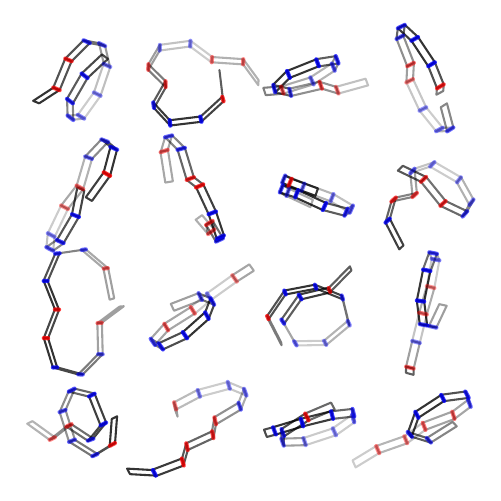}
    \caption{Sample ladder polymer configurations generated using MC with $(L,R_f,\alpha)=(12,0.14,0.89)$ and $K_t=K_b=100$.}
    \label{fig: exp_config}
\end{figure}

Tab.~\ref{tab: ML_polymer_parameter} shows the four GPR predicted polymer parameters of SANS our synthesized CANAL ladder polymer along with comparison with parameters obtained from other traditional methods. Note that our ML inversion method can extract all parameters simultaneously, and those parameters that the traditional method can extract, $(\alpha, L, R_g^2)$, show strong agreement with its results. Moreover, due to the special monomer structure of the CANAL ladder polymer, the anti rate $R_a$ is a unique parameter that only can be obtained using the ML inversion method. The ML inversion method suggest our sample is relative short, with only about 12 segments, and the radius of gyration is even just $R_g^2\simeq 2$, fairly small for such contour length comparing to semiflexible chains\cite{ding2024off}. This discrepancy is explained by the low anti rate $R_a\simeq0.14$, which suggesting the monomers are most connected through Sync link, making the polymer roll up, as shown in Fig.~\ref{fig: exp_config}. The tendency to have more coiling structure of ladder polymers has also been observed from other systems.\cite{ikai2019triptycene}

\section{Conclusions}
In this paper, we introduce an biaxial segmentation model for the ladder polymer, we use a ML inversion method to extract polymer parameters from the form factor of the polymer system, and apply such method on real scattering data of the CANAL ladder polymer. The segmentation model represent the ladder polymer as a chain of two dimensional rectangular segments whose orientation is given by two unit vectors corresponding to the long and short axis. We prepare a data set consisting of 6,000 form factor $\vb{F}=\left\{ S(QB)\right\}$ and corresponding polymer parameters $\vb{Y}=\left\{(R_a,\alpha,L,R_g^2)\right\}$ including anti rate $R_a$, inherent bending angle $\alpha$, contour length $L$, and radius of gyration $R_g^2$. We train a GPR using part of data set as training set to achieve the mapping from $\vb{F}$ to $\vb{Y}$, and show that the trained GPR achieves excellent mapping when applied on the rest of data set, i.e. test set. Given that, we apply the ML inversion analysis on real scattering data. We firstly synthesize a CANAL ladder polymer, and run SANS experiment for a dilute sample. We normalize the SANS measured $S(SQ)$ and feed it into the trained GPR. All four polymer parameters are successfully extracted and the consistent with other traditional method when applicable. The anti rate $R_a$ is extracted from the scattering data for the first time, providing new insight for the understanding of ladder polymer.

Using the ML extracted polymer parameter, we can regenerate sample configurations using MC, it is expected that the CANAL ladder polymer sample we synthesized roll up to a coil or ring shape due to it's low anti rate. Further studies on single polymer imaging using scanning tunneling microscope\cite{binnig1987scanning} (STM) or ultra resolution atomic force microscopy\cite{giessibl2003advances} (AFM) would be highly beneficial. Moreover, the sample we used in this work only have inherent bending, application of this ML inversion method for other CANAL ladder polymer with both inherent bending and twisting can also be carried out in the future.  

We also note that the CANAL ladder polymer structure we studied is dominated by the inherent bending angle and anti rate, the effect of bending modulus $K_b$ and twisting modulus $K_t$ are too weak to be extracted for this system. For the study of these $K_b$ and $K_t$, ladder polymer whose monomers are connected in a flat manner are more suitable, as well as conjugated polymer\cite{yin2022synthesis,cao2023molecular} whose twisting can be more significant due to the existent of single bond.

\begin{acknowledgments}
We thank Jihua Chen and Dale Hensley for fruitful discussions and exploratory scanning electron microscope experiment. This research was performed at the Spallation Neutron Source and the Center for Nanophase Materials Sciences, which are DOE Office of Science User Facilities operated by Oak Ridge National Laboratory. This research was sponsored by the Laboratory Directed Research and Development Program of Oak Ridge National Laboratory, managed by UT-Battelle, LLC, for the U. S. Department of Energy. The ML aspects were supported by by the U.S. Department of Energy Office of Science, Office of Basic Energy Sciences Data, Artificial Intelligence and Machine Learning at DOE Scientific User Facilities Program under Award Number 34532. Monte Carlo simulations and computations used resources of the Oak Ridge Leadership Computing Facility, which is supported by the DOE Office of Science under Contract DE-AC05-00OR22725 and resources of the National Energy Research Scientific Computing Center, which is supported by the Office of Science of the U.S. Department of Energy under Contract No. DE-AC02-05CH11231.
\end{acknowledgments}



\bibliography{apssamp}

\pagebreak
\widetext
\begin{center}
\textbf{\large Supporting Information}
\end{center}
\setcounter{equation}{0}
\setcounter{figure}{0}
\setcounter{table}{0}
\setcounter{page}{1}
\makeatletter
\renewcommand{\theequation}{S\arabic{equation}}
\renewcommand{\thefigure}{S\arabic{figure}}
\renewcommand{\bibnumfmt}[1]{[S#1]}
\renewcommand{\citenumfont}[1]{S#1}

\textbf{Characterization of synthesized ladder polymer} All chemicals, including 1,4-diethylbenzene, were purchased from commercial sources and used as received unless otherwise noted. 1,4-dibromo-2,5-diethylbenzene was prepared following known literature procedures (Sugamata et al., 2021). All reactions were performed under nitrogen in oven- or flame-dried glassware unless otherwise noted. Flash column chromatography was carried out with Silica 60 (230--400 mesh; Fisher). Analytical thin-layer chromatography (TLC) was carried out using 0.2 mm silica gel plates (silica gel 60, F254, EMD Chemical).

Proton nuclear magnetic resonance (\(^{1}\mathrm{H}\) NMR) and carbon nuclear magnetic resonance (\(^{13}\mathrm{C}\) NMR) spectra were recorded in \(\mathrm{CDCl}_3\) using 500 MHz Varian NMR spectrometers. Chemical shifts are reported in parts per million (ppm) relative to residual protonated solvent for \(^{1}\mathrm{H}\) (\(\mathrm{CHCl}_3 = \delta 7.26\), \(\mathrm{DCM} = \delta 5.30\), \(\mathrm{methanol}\text{-}\mathrm{d}_4 = \delta 3.31\)) and relative to carbon resonances of the solvent for \(^{13}\mathrm{C}\) (\(\mathrm{CDCl}_3 = \delta 77.0\)). Data are reported as follows: chemical shift, multiplicity (s = singlet, d = doublet, dd = doublet of doublets, t = triplet, q = quartet, m = multiplet, br = broad signal, and associated combinations), coupling constant(s) (Hz), and integration.

\begin{figure}[h]
    \centering
    \includegraphics[width=1\linewidth]{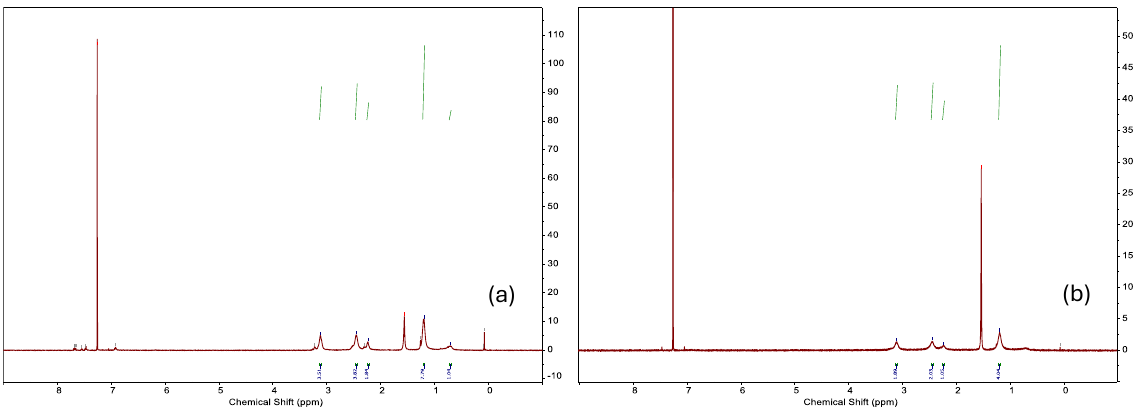}
    \caption{Nuclear magnetic resonance spectra for the CANAL ladder polymer. (a) Low molecular weight. (b) high molecular weight}
    \label{fig: NMR}
\end{figure}

\begin{figure}[h]
    \centering
    \includegraphics[width=0.5\linewidth]{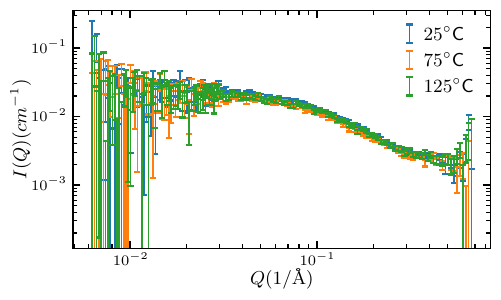}
    \caption{Scattering intensity of the synthesized CANAL polymer under three different temperature: \SI{25}{\celsius}, \SI{75}{\celsius} and \SI{125}{\celsius}.}
    \label{fig: exp_Sq_temp}
\end{figure}

\textbf{Effect of temperature variation} We confirm the sample is fully dissolved and without aggregation by comparing the scattering function $I(Q)$ measured under different temperatures: \SI{25}{\celsius}, \SI{75}{\celsius} and \SI{125}{\celsius}. As shown in Fig.~\ref{fig: exp_Sq_temp}, the $I(Q)$ spectrum are consistent cross all three temperatures, indicating that the polymer sample we used is fully dissolved.

\end{document}